\documentclass{article}
\usepackage{amsfonts}

\newcommand{\beqn}{\begin{eqnarray}}
\newcommand{\eeqn}{\end{eqnarray}}
\newcommand{\rd}{\partial}

\newcommand{\Res}{\mathop{{\mathrm{Res}}}}
\newcommand{\diag}{\mathop{{\mathrm{diag}}}}

\newcommand{\Jac}{{\mathrm{Jac}}}
\newcommand{\Prym}{{\mathrm{Prym}}}
\newcommand{\SW}{{\mathrm{SW}}}
\newcommand{\KdV}{{\mathrm{KdV}}}

\newcommand{\bbC}{{\mathbb{C}}}

\newcommand{\bbZ}{{\mathbb{Z}}}
\newcommand{\frakg}{{\mathfrak{g}}}

\newcommand{\calF}{{\mathcal{F}}}
\newcommand{\calH}{{\mathcal{H}}}

\newcommand{\calO}{{\mathcal{O}}}
\newcommand{\calP}{{\mathcal{P}}}
\newcommand{\calT}{{\mathcal{T}}}
\newcommand{\calU}{{\mathcal{U}}}
\newcommand{\Omegahat}{{\hat{\Omega}}}
\newcommand{\tauhat}{{\hat{\tau}}}
\newcommand{\Xhat}{{\hat{X}}}
\newcommand{\abar}{{\bar{a}}}
\begin{document}

\title{Whitham Deformations and Tau Functions 
in $N = 2$ Supersymmetric Gauge Theories
\footnote{Talk given at the workshop 
``Gauge Theory and Integrable Models'', 
YITP, Kyoto, January 26-29, 1999}} 

\author{Kanehisa Takasaki\\
{\normalsize Department of Fundamental Sciences, Kyoto University}\\
{\normalsize Yoshida, Sakyo-ku, Kyoto 606-8501, Japan}\\
{\normalsize E-mail: takasaki@yukawa.kyoto-u.ac.jp}}

\date{KUCP-0136, hep-th/9905224}

\maketitle

\begin{abstract} 
We review new aspects of integrable systems discovered recently 
in $N=2$ supersymmetric gauge theories and their topologically 
twisted versions.  The main topics are (i) an explicit construction 
of Whitham deformations of the Seiberg-Witten curves for classical 
gauge groups, (ii) its application to contact terms in the $u$-plane 
integral of topologically twisted theories, and (iii) a connection 
between the tau functions and the blowup formula in topologically 
twisted theories.  
\end{abstract}

\section{Introduction}
\setcounter{equation}{0}

The Seiberg-Witten low energy effective action of 
four-dimensional $N = 2$ supersymmetric gauge 
theories (with and without matters) \cite{bib:Se-Wi94} 
is described by the geometry of a family of complex 
algebraic curves (the Seiberg-Witten curves) $C$ 
fibered over the Coulomb moduli space $\calU$. 
Each curve is equipped with a meromorphic 
differential (the Seiberg-Witten differential) 
$dS_\SW$.  This differential induces a ``special 
geometry'' on $\calU$ with special coordinates $a_j$ 
and their duals $a^D_j$ defined by the period 
integrals of $dS_\SW$ along suitable cycles 
$\alpha_j$ and $\beta_j$ on $C$.  The prepotential 
$\calF$ of this special geometry determines the 
effective action.   

Shortly after the original $SU(2)$ theories of Seiberg 
and Witten were generalized to other gauge groups 
\cite{bib:KKTS94,bib:Ar-Fa94,bib:Da-Su95,
bib:Ha-Oz95,bib:Br-La95,bib:Ar-Sh95,bib:Hanany95}, 
a close connection with integrable systems was 
discovered \cite{bib:GKMMM95,bib:Ma-Wa95},  
and further pursued 
\cite{bib:Na-Ta95,bib:Do-Wi95,bib:Eg-Ya95,
bib:Martinec95,bib:Go-Ma95,bib:It-Mo95,
bib:GMMM96,bib:Marshakov96,bib:Ah-Na96,bib:Kr-Ph96,
bib:Nekrasov96,bib:DH-Kr-Ph96,bib:Gorsky96,
bib:Marshakov97,bib:Ma-Ma-Mo97,bib:DH-Ph97,
bib:Kapustin98,bib:DH-Ph98} 
as a guiding principle for the study of 
$N = 2$ supersymmetric gauge theories.  
This connection appears in two different aspects. 
The first aspect is a direct relation with 
integrable systems such as the periodic Toda 
systems and the elliptic Calogero-Moser systems.  
In this connection, the Seiberg-Witten curves 
are interpreted to be the spectral curves of 
a finite-dimensional integrable system, the 
special coordinates are nothing but the action 
variables, and the electro-magnetic duality 
of the abelianized theory is encoded into the 
structure of an associated polarized Abelian 
variety.  Thus the building blocks of the 
Seiberg-Witten effective theory fit well to 
the general setup of integrable systems, in 
particular, of ``algebraically integrable 
systems'' \cite{bib:Donagi97,bib:Freed97}.  
The second aspect is the notion of Whitham 
deformations of the spectral curve.  
This leads to a new interpretation of 
recursion relations of instanton expansion 
\cite{bib:Matone95} and the so called 
RG equation \cite{bib:So-Th-Ya96,bib:Bo-Ma96}. 

In this paper, we review recent developments 
in these lines 
\cite{bib:GMMM98,bib:Takasaki98,bib:Ed-Ma-Ma98,
bib:Ed-Ma99,bib:Takasaki99,bib:Ed-GR-Ma99,
bib:DH-Ph99,bib:Marshakov99,bib:Marino99}, 
mostly focussed on the connection with 
topologically twisted $N = 2$ theories 
\cite{bib:Mo-Wi97,bib:Lo-Ne-Sh97-98,bib:Ma-Mo97-98}. 
Section 2 is a short summary of the fundamental 
geometric stuff (complex algebraic curves, 
differentials, special coordinates and prepotentials).  
Section 3 is concerned with an explicit construction 
of Whitham deformations, which has been applied to 
instanton expansion and soft breaking of $N = 2$ 
supersymmetry, as well as the topological theories.  
Section 4 deals with the relation to the topological
theories in more detail. Here a central role is 
played by the notion of tau functions.

\section{Seiberg-Witten Geometry for Classical Gauge Groups}
\setcounter{equation}{0}

\subsection{Seiberg-Witten geometry for $SU(\ell+1)$}

The Seiberg-Witten curve $C$ for the gauge group 
$SU(\ell+1)$ is a complex algebraic curve of the form 
\cite{bib:GKMMM95} 
\beqn
    z + \frac{\mu^2}{z} = P(z) 
    = x^{\ell+1} - \sum_{j=2}^{\ell+1} u_j x^{\ell+1-j}, 
\eeqn
where $\mu$ denotes the power $\Lambda^{\ell+1}$ of 
the renormalization group parameter $\Lambda$, and 
$u_j$'s are the Coulomb moduli.  This is a hyperelliptic 
curve of genus $\ell$.   By the simple transformation 
\beqn
    z = \frac{P(x) + y}{2},  
\eeqn
the above equation can be indeed converted to the 
usual hyperelliptic equation 
\beqn
    y^2 = P(z)^2 - 4\mu^2.  
\eeqn

The prepotential of the Seiberg-Witten effective 
theory is defined by use of special coordinates 
$a_j,a^D_j$ ($j=1,\cdots,\ell$) on the Coulomb 
moduli space $\calU$.  These special coordinates 
are given by the period integrals
\footnote{As opposed to my previous papers
\cite{bib:Takasaki98,bib:Takasaki99}, 
I have inserted the factor ``$1/(2\pi i)$'', 
which is rather standard in the physical 
literature.  Similar changes of convention 
have been done throughout this paper.} 
\beqn
    a_j = \frac{1}{2\pi i}\oint_{\alpha_j} dS_\SW, 
    \quad 
    a^D_j = \frac{1}{2\pi i}\oint_{\beta_j} dS_\SW, 
\eeqn
where $\alpha_j$ and $\beta_j$ are a symplectic 
basis of cycles on the curve $C$, and $dS_\SW$ 
the Seiberg-Witten differential 
\beqn
    dS_\SW = x \frac{dz}{z}. 
\eeqn
The integrals $a_j$ and $a^D_j$ give, respectively, 
a set of local coordinates on the Coulomb moduli 
space.  If one interpret $a_j$ as such coordinates 
and the duals $a^D_j$ as a function of the former, 
one can define the prepotential 
$\calF = \calF(a_1,\cdots,a_\ell)$ as a solution 
of the equations 
\beqn
    \frac{\rd \calF}{\rd a_j} = a^D_j. 
\eeqn

Furthermore, the second derivatives of $\calF$ 
reproduce the period matrix 
$\calT = (\calT_{jk})_{j,k=1,\cdots,\ell}$: 
\beqn
    \frac{\rd^2\calF}{\rd a_j \rd a_k} 
    = \calT_{jk} 
    = \frac{1}{2\pi i}\oint_{\beta_j} d\omega_k. 
\eeqn
Here $d\omega_j$ ($j = 1,\cdots,\ell$) are a basis 
of holomorphic differentials on $C$ normalized as 
\beqn
    \frac{1}{2\pi i}\oint_{\alpha_j} d\omega_k 
    = \delta_{jk}. 
\eeqn
This matrix $\calT$ appears in the analytic expression 
\beqn
    \Jac(C) \simeq \bbC^\ell/(\bbZ^\ell + \calT \bbZ^\ell) 
\eeqn
of the  Jacobi variety $\Jac(C)$ of the curve $C$.  

In the context of integrable systems, the curve $C$ 
is nothing but the spectral curve $\det(L(z) - xI) = 0$ 
of a Lax representation of the $\ell$-periodic Toda 
system, i.e., the affine Toda system of the 
$A^{(1)}_\ell$ type; the Lax pair is constructed 
in the vector representation of $sl(\ell+1)$.  
It is well known \cite{bib:Da-Ta76,bib:Krichever78,
bib:vM-Mu79,bib:Ad-vM80} that the dynamics of this 
Toda system is mapped to linear flows on the Jacobi 
variety $\Jac(C)$. In a more abstract language, 
the $2\ell$-dimensional total space of the fiber 
bundle of the Jacobi varieties over the Coulomb 
moduli space $\calU$ becomes an algebraically 
integrable system \cite{bib:Donagi97,bib:Freed97}.

\subsection{Seiberg-Witten curves for other classical 
gauge groups} 

The Seiberg-Witten curves for other classical gauge 
groups (i.e., orthogonal and symplectic groups) 
can be written 
\beqn
    z + \frac{\mu^2}{z} = W(x), 
\eeqn
where $W(x)$ is a polynomial or a Laurent polynomial 
of the form 
\beqn
    SO(2\ell+1) &:& W(x) = x^{-1} \left(x^{2\ell} 
      - \sum_{j=1}^\ell u_j x^{2\ell-2j} \right). 
    \nonumber \\
    Sp(2\ell) &:&  W(x) = x^2 \left(x^{2\ell} 
      - \sum_{j=1}^\ell u_j x^{2\ell-2j} \right) + 2\mu. 
    \nonumber \\
    SO(2\ell) &:& W(x) = x^{-2} \left(x^{2\ell} 
      - \sum_{j=1}^\ell u_j x^{2\ell-2j} \right). 
\eeqn
In the case of $SU(\ell+1)$ and $SO(2\ell)$, 
$W(x)$ coincide with the superpotential of the 
topological Landau-Ginzburg models (or $d < 1$ 
strings) for the $A_\ell$ and $D_\ell$ isolated 
singularities \cite{bib:DVV91,bib:Bl-Va92}.  
Guided by this analogy, basic notions in the 
topological Landau-Ginzburg models, such as the 
Gauss-Manin system and the WDVV equations, have 
been generalized to the Seiberg-Witten effective 
theories \cite{bib:Bo-Ma-WDVV, bib:Ma-Mi-Mo-WDVV,
bib:Ma-Mo-WDVV,bib:Be-Ma-WDVV,bib:It-Ya-GM,
bib:It-Xi-Ya-GM}. 

These curves coincide with the spectral curve 
$\det(L(z) - xI) = 0$ of a Lax pair (in minimal 
dimensions) of an affine Toda system, namely, 
the affine Toda system associated with the dual 
Lie algebra $\frakg^{(1)\vee}$ of the affine 
algebra $\frakg^{(1)}$ of the gauge group $G$ 
\cite{bib:Ma-Wa95}. Note that the dual Lie 
algebra $\frakg^{(1)\vee}$ for the non-simply 
laced gauge groups $SO(2\ell+1)$ and $Sp(2\ell)$ 
is a twisted affine algebra. 

All these curves are hyperelliptic.  One can find 
an equivalent expression in the usual expression 
$y^2 = R(x)$ of hyperelliptic curves: 
\beqn 
    SO(2\ell+1) &:& 
      y^2 = Q(x^2)^2 - 4 \mu^2 x^2, \ 
      z = \bigl(Q(x^2) + y \bigr) / 2x. 
      \nonumber \\
    Sp(2\ell) &:& 
      y^2 = Q(x^2) \bigl(x^2 Q(x^2) + 4 \mu \bigr), \ 
      z = \bigl(2 \mu + x^2 Q(x^2) + xy \bigr) / 2. 
      \nonumber \\
    SO(2\ell) &:& 
      y^2 = Q(x^2)^2 - 4 \mu^2 x^4, \ 
      z = \bigl( Q(x^2) + y \bigr) / 2. 
      \nonumber 
\eeqn
Actually, it is in this form (or the quotient curve $C'$ 
discussed later on) that the relevant complex algebraic
curves for classical gauge groups were first derived 
\cite{bib:KKTS94,bib:Ar-Fa94,bib:Da-Su95,bib:Ha-Oz95,
bib:Br-La95,bib:Ar-Sh95,bib:Hanany95}.

\subsection{Involutions and Prym varieties}

The curves $C$ for the orthogonal and symplectic gauge 
groups have two involutions: 
\beqn
    \sigma_1 &:& 
      (x,y) \mapsto (x,-y), \quad  
      (x,z) \mapsto (x, \mu^2/z), 
    \nonumber \\
    \sigma_2 &:& 
      (x,y) \mapsto (-x,y), \quad 
      (x,z) \mapsto (-x,z). 
\eeqn
The first involution is the hyperelliptic involution, 
which also exists on the curve for $SU(\ell+1)$; the 
second involution is a characteristic of the other cases. 
The quotient $C_2 = C / \sigma_2$ by the second involution 
is again a hyperelliptic curve.  Some fundamental properties 
of these curves are summarized in the following table. 

\begin{center}
\begin{tabular}{|c|c|c|c|}
\hline
$G$ & ${\mathrm{genus}}(C)$ & ${\mathrm{genus}}(C_2)$ & 
  covering $C \to C_2$ \\
\hline \hline 
$Sp(2\ell)$ & $2\ell$ & $\ell$ & unramified \\
\hline
$SO(2\ell+1),SO(2\ell)$ & $2\ell-1$ & $\ell-1$ & ramified \\
\hline
\end{tabular}
\end{center}

The double covering $C \to C_2$ determines the Prym 
variety $\Prym(C/C_2)$.  This is an $\ell$-dimensional 
polarized Abelian variety, which plays the role of the 
Jacobi variety $\Jac(C)$ for the $SU(\ell+1)$ gauge 
theory. Let us specify the structure of this Prym 
variety in some detail \cite{bib:Fay73}.   

In the complex analytic language, $\Prym(C/C_2)$ is 
a complex torus of the form 
\beqn
    \Prym(C/C_2) \simeq 
    \bbC^\ell / (\Delta \bbZ^\ell + 2 \calP \bbZ^\ell), 
\eeqn
where $\Delta$ is a diagonal matrix 
$\Delta = \diag(d_1,\cdots,d_N)$ with positive integers 
on the diagonal line, and $\calP$ is a complex symmetric 
matrix $(\calP_{jk})$ with positive definite imaginary 
part. The diagonal elements of $\Delta$ represent 
the polarization: 
\beqn
    Sp(2\ell) &:& 
      (d_1,\cdots,d_\ell) = (2,\cdots,2,2), 
    \nonumber \\
    SO(2\ell+1), SO(2\ell) &:& 
      (d_1,\cdots,d_\ell) = (2,\cdots,2,1). 
\eeqn
The matrix elements of $\calP = (\calP_{jk})$ are given 
by the period integrals 
\beqn
    \calP_{jk} = \frac{d_j}{4\pi i} \oint_{\beta_j} d\omega_k, 
\eeqn
where $d\omega_j$ ($j = 1,\cdots,\ell$) are holomorphic 
differentials on $C$ that are ``odd'' under the action of 
$\sigma_2$, i.e., 
\beqn
    \sigma_2^* d\omega_j = - d\omega_j, 
\eeqn
and uniquely determined by the normalization condition 
\beqn
    \frac{1}{2\pi i}\oint_{\alpha_j} d\omega_k = \delta_{jk}. 
\eeqn
The cycles $\alpha_j,\beta_j$ ($j = 1,\cdots,\ell$) in 
these period integrals have to be chosen as follows: 
\begin{itemize}
\item 
For $Sp(2\ell)$: 
The $4\ell$ cycles $\alpha_j,-\sigma_2(\alpha_j),
\beta_j,-\sigma_2(\beta_j)$ ($j = 1,\cdots,\ell$) form 
a symplectic basis of cycles of $C$. 
\item 
For $SO(2\ell+1)$ and $SO(2\ell)$: 
The homology classes $[\alpha_\ell]$ and $[\beta_\ell]$ 
are ``odd'' under the action of $\sigma_2$, i.e., 
$\sigma_2([\alpha_\ell]) = - [\alpha_\ell]$ and 
$\sigma_2([\beta_\ell]) = - [\beta_\ell]$.  
The $4\ell - 2$ cycles $\alpha_j,-\sigma_2(\alpha_j),
\beta_j,-\sigma_2(\beta_j)$ ($j = 1,\cdots,\ell-1$) 
and $\alpha_\ell,\beta_\ell$ altogether form 
a symplectic basis of cycles of $C$. 
\end{itemize}
In particular, these cycles have the intersection 
numbers $\alpha_j \cdot \beta_k = \delta_{jk}$, 
$\alpha_j \cdot \alpha_k = \beta_j \cdot \beta_k = 0$.

\subsection{Special coordinates and prepotential} 

The Seiberg-Witten differential for the orthogonal 
and symplectic gauge groups is given by 
\beqn
    dS_{\SW} = x \frac{dz}{z} 
             = \frac{xW'(x)dx}{\sqrt{W(x)^2 - 4\mu^2}}. 
\eeqn
This differential, like $d\omega_j$, is odd'' under 
the action of $\sigma_2$: 
\beqn
    \sigma_2^* dS_{\SW} = - dS_{\SW}. 
\eeqn

Given a set of cycles $\alpha_j,\beta_j$ as mentioned 
above, one can define the special coordinates 
$a_j$ and their duals $a^D_j$ on the Coulomb moduli 
space $\calU$ by the period integrals 
\beqn
    a_j = \frac{1}{2\pi i}\oint_{\alpha_j} dS_{\SW}, 
    \quad 
    a_j^D = \frac{1}{2\pi i}\oint_{\beta_j} dS_{\SW}. 
\eeqn
The prepotential $\calF = \calF(a_1,\cdots,a_\ell)$ 
is again characterized by the differential equation 
\beqn
    \frac{\rd \calF}{\rd a_j} = a_j^D. 
\eeqn
The matrix elements $\calP_{jk}$ of $\calP$ can be 
expressed as second derivatives of the prepotential:
\beqn
    \frac{\rd^2 \calF}{\rd a_j \rd a_k} = \calP_{jk}. 
\eeqn

\subsection{Quotient curve of genus $\ell$} 

The Prym variety $\Prym(C/C_2)$ can be identified, 
up to isogeny, with the Jacobi variety $\Jac(C')$ 
of the quotient curve $C' = C/\sigma'$ obtained by 
the following involution: 
\beqn
    Sp(2\ell) &:& \sigma' = \sigma_2, 
    \nonumber \\
    SO(2\ell+1), SO(2\ell) &:& \sigma' = \sigma_1 \sigma_2. 
\eeqn

The quotient curve $C'$ is also hyperelliptic and has 
genus $\ell$.  The holomorphic differentials $d\omega_j$,  
as well as $dS_\SW$, are ``even'' (i.e., invariant) under 
the action of $\sigma'$, so that they are the pull-back 
of differentials on $C'$.  The matrix $\calP$ is actually 
the period matrix of $C'$: 
\beqn
    \Jac(C') \simeq \bbC^\ell/(\bbZ^\ell + \calP \bbZ^\ell). 
\eeqn

To find an explicit expression of $C'$, one can use 
the following invariants $\xi$ and $\eta$ of $\sigma'$: 
\beqn
    Sp(2\ell) &:& \xi = x^2, \quad  \eta = y, 
    \nonumber \\
    SO(2\ell+1), SO(2\ell) &:& \xi = x^2, \quad \eta = xy. 
\eeqn
In terms of these coordinates, $C'$ can be written 
as follows: 
\beqn
    SO(2\ell+1) &:& 
      \eta^2 = \xi \Bigl( Q(\xi^2) - 4 \mu^2 \xi \Bigr). 
      \nonumber \\
    Sp(2\ell) &:& 
      \eta^2 = Q(\xi) \Bigl( \xi Q(\xi) + 4 \mu \Bigr). 
      \nonumber \\
    SO(2\ell) &:& 
      \eta^2 = \xi \Bigl( Q(\xi)^2 - 4 \mu^2 \xi^2 \Bigr). 
\eeqn

Let us compare the two curves $C$ and $C'$.  The curve $C$ 
is a double covering of the $x$-sphere and has two points 
$P_\infty^\pm$ at infinity above $x = \infty$.  These two 
points correspond to $z = \infty$ and $z = 0$, and mapped 
to each other by the hyperelliptic involution $\sigma_1$.  
This is a characteristic of the spectral curves of affine 
Toda systems \cite{bib:Da-Ta76,bib:Krichever78,bib:vM-Mu79,
bib:Ad-vM80}. 
The curve $C'$, in contrast, has a single point at infinity 
above the $\xi$-shere, so that $C'$ is branched over 
$\xi = \infty$.  Hyperelliptic curves of this type arise 
in the KdV  hierarchy \cite{bib:Da-Ta76,bib:Du-Ma-No76,
bib:Krichever77}.  
It is well known that the KdV hierarchy is a specialization 
the KP hierarchy \cite{bib:Sa-Sa82,bib:DJKM83,bib:Se-Wi85}, 
in which only ``odd'' time variables $t_{2n-1}$ remain 
non-trivial among all possible flows $t_1,t_2,t_3,\cdots$ 
of the KP hierarchy.  We shall see a similar structure 
later on in Whitham deformations.

\section{Construction of Whitham Deformations} 
\setcounter{equation}{0}

\subsection{What are Whitham deformations?}

In all the aforementioned cases, the Seiberg-Witten 
differential $dS_\SW$ plays the role of a ``generating 
differential'', namely, differentiating against the 
moduli $u_j$ give a basis of (odd) holomorphic 
differentials: 
\beqn
    \left.\frac{\rd}{\rd u_j}dS
      \right|_{z={\mathrm{const.}}} 
    = dv_j. 
\eeqn
Here ``$(\cdots)|_{z={\mathrm{const.}}}$'' means 
differentiating while leaving $z$ constant.  
For instance, in the case of $SU(\ell+1)$, the 
following  holomorphic differentials $dv_j$ are 
thus reproduced: 
\beqn
    dv_j = \frac{x^{\ell+1-j}dx}{y} \quad 
    (j = 2,\cdots,\ell+1). 
\eeqn

If we use $a_j$ as independent variables, the 
outcome are the normalized holomorphic 
differentials: 
\beqn
    \left.\frac{\rd}{\rd a_j}dS
      \right|_{z={\mathrm{const.}}} 
    = d\omega_j. 
\eeqn
Note that the moduli $u_j$ in this situation are 
a function $u_j = u_j(\vec{a})$ of $\vec{a} = 
(a_1,\cdots,a_\ell)$, which gives an inverse of 
the period map $\vec{u} \mapsto \vec{a}$, 
$\vec{u} = (u_2,\cdots,u_{\ell+1})$.  The curve $C$ 
is now deformed as $\vec{a}$ varies, $C = C(\vec{a})$. 

``Whitham deformations'' are an extension of these 
deformations with new ``time variables'' $T_n$ 
($n = 1,2,\cdots$). More precisely, we consider the 
following setup.  

\begin{itemize} 
\item 
The moduli $u_j$ of the curve $C$ are deformed as 
a function $u_j = u_j(\vec{a},\vec{T})$ of $\vec{a}$ 
and $\vec{T} = (T_1,T_2,\cdots)$.  This induces a 
family of deformations $C \mapsto C(\vec{a},\vec{T})$ 
of the curve $C$. At $\vec{T} = (1,0,0,\cdots)$, they 
are required to reduce to the Seiberg-Witten family.  
It should be noted that, apart from this 
``Seiberg-Witten point'', the $a_j$'s are 
{\it no longer identical} to the special coordinates 
for the deformed curve $C(\vec{a},\vec{T})$ defined 
by the period integrals of the Seiberg-Witten 
differential on this curve.  

\item
The following equations are satisfied: 
\beqn
    \left.\frac{\rd}{\rd a_j}dS
      \right|_{z={\mathrm{const.}}} 
    = d\omega_j, 
    \quad 
    \left.\frac{\rd}{\rd T_n}dS
      \right|_{z={\mathrm{const.}}} 
    = d\Omega_n. 
\eeqn
Here $d\Omega_n$ ($n =1,2,\cdots$) are meromorphic 
differentials of the second kind with poles at 
$P_\infty^\pm$ only (with some more conditions on 
the singular part, see below) and vanishing 
$\alpha$-periods 
\beqn
    \oint_{\alpha_j} d\Omega_n = 0 
    \quad (j = 1,\cdots,\ell), 
\eeqn
and $dS$ is a linear combination of these meromorphic 
differentials and the normalized holomorphic 
differentials of the form 
\beqn
    dS = \sum_{n\ge 1} T_n d\Omega_n 
       + \sum_{j=1}^N a_j d\omega_j. 
\eeqn

\item
$dS$ reduces to $dS_\SW$ at $\vec{T} = (1,0,0,\cdots)$. 

\item
For the prepotential $\calF$ to be extended to this 
family of deformations, the singular behavior of 
$d\Omega_n$ at $P_\infty^\pm$ (i.e., $z = \infty$ 
and $z = 0$) should be of the form 
\beqn
    d\Omega_n = df_n^\pm(z) + \mbox{non-singular}, 
\eeqn
where $f_n^\pm(z)$ is a polynomial in $z^\pm$ with 
{\it constant} coefficients.  
\end{itemize}

\subsection{Construction of Whitham deformations 
for $SU(\ell+1)$} 

The solution of Gorsky et al. \cite{bib:GMMM98} for 
the $SU(\ell+1)$ gauge theory is constructed by the 
following steps. 

\begin{enumerate} 
\item 
Consider the meromorphic differentials 
\beqn
    d\Omegahat_n = R_n(x) \frac{dz}{z}, \quad 
    R_n(x) = \Bigl( P(x)^{n/(\ell+1)} \Bigr)_{+}. 
\eeqn
Here $(\cdots)_{+}$ denotes the polynomial part of 
a Laurent series of $x$.  The fractional power of 
$P(x)$ is understood to be a Laurent series of 
the form $x^n + \cdots$ at $x = \infty$. Since 
$R_1(x) = x$, $d\Omegahat_1$ is nothing but the 
Seiberg-Witten differential.  
\item 
Consider the differential 
\beqn
    dS = \sum_{n\ge 1} T_n d\Omegahat_n 
\eeqn
and its period integrals 
\beqn
    a_j = \frac{1}{2\pi i} \oint_{\alpha_j} dS 
    = \sum_{n\ge 1} \frac{T_n}{2\pi i} 
      \oint_{\alpha_j}d\Omegahat_n. 
\eeqn
These period integrals are functions of the moduli 
$u_j$ and the deformation parameters $T_n$.  
They determine a family of deformations of the 
Seiberg-Witten period map $\vec{u} \mapsto \vec{a}$ 
with parametes $T_n$.  
\item 
The period map $\vec{u} \mapsto \vec{a}$ from 
the $\vec{u}$-space to the $\vec{a}$-space is 
invertible if $\vec{T}$ is close to $(1,0,0,\cdots)$, 
because the Seiberg-Witten period map at this 
point is invertible.  The inverse map $\vec{a} 
\mapsto \vec{u} = \Bigl(u_2(\vec{a},\vec{T}),
\cdots,u_{\ell+1}(\vec{a},\vec{T})\Bigr)$ 
gives deformations of the Seiberg-Witten moduli 
$u_j = u_j(\vec{a})$, hence of the curve $C$, 
with parameters $T_n$.  
\item 
The differentials 
\beqn
    d\Omega_n = d\Omegahat_n 
      - \sum_{j=1}^N c_j^{(n)} d\omega_j, \quad 
    c_j^{(n)} = 
      \frac{1}{2\pi i} \oint_{\alpha_j} d\Omegahat_n 
\eeqn
satisfy the required normalization condition. 
$dS$ thereby becomes a linear combination of $d\Omega_n$ 
and $d\omega_j$ of the required form. 
\end{enumerate}
The following property of $d\Omegahat_n$ plays a 
central role in this construction: 
\begin{itemize}
\item $(\rd/\rd u_j)d\Omegahat_n |_{z={\mathrm{const.}}}$ 
are holomorphic differentials on $C$. 
\end{itemize}
Note that this is a generalization of the property 
of the Seiberg-Witten differential that we have 
mentioned in the beginning of this section. 
Once this property is established, it is rather 
straightforward to verify that the other 
requirements are indeed fulfilled.  

Somewhat delicate part is the determination of 
the Laurent polynomials $f_n^\pm(z)$ that 
represent the singular part of $d\Omega_n$ at 
$P_\infty^\pm$.  This can be worked out by using 
the identity 
\beqn
    P(x)^{n/(\ell+1)} 
    = \Bigl(z + \frac{\mu^2}{z}\Bigr)^{n/(\ell+1)}. 
\eeqn
The singular part of Laurent expansion of the right 
hand side at $z = \infty$ or $z = 0$ determines 
the Laurent polynomials $f_n^\pm(z)$. 
Obviously the singular part is a Laurent polynomial 
with constant coefficients.  Accordingly $f_n^\pm(z)$, 
too, turn out to have constant coefficients. 

Having this solution, we can now define the prepotential 
$\calF = \calF(\vec{a},\vec{T})$ by the equation 
\beqn
    \frac{\rd \calF}{\rd a_j} &=& 
    \frac{1}{2\pi i} \oint_{\beta_j} dS, 
    \nonumber \\
    \frac{\rd \calF}{\rd T_n} &=& 
    - \frac{1}{(2\pi i)^2}\oint_{P_\infty^+} f_n^+(z) dS 
    - \frac{1}{(2\pi i)^2}\oint_{P_\infty^-} f_n^-(z) dS. 
\eeqn
The second derivatives are also related to period 
integrals: 
\beqn
    \frac{\rd^2 \calF}{\rd a_j \rd a_k} &=& 
    \frac{1}{2\pi i} \oint_{\beta_j} d\omega_k, 
    \nonumber \\
    \frac{\rd^2 \calF}{\rd a_j \rd T_n} &=& 
    - \frac{1}{(2\pi i)^2}\oint_{P_\infty^+} 
        f_n^+(z) d\omega_j 
    - \frac{1}{(2\pi i)^2}\oint_{P_\infty^-} 
        f_n^-(z) d\omega_j, 
    \nonumber \\
    \frac{\rd^2 \calF}{\rd T_m \rd T_n} &=& 
    - \frac{1}{(2\pi i)^2} \oint_{P_\infty^+} 
        f_m^+(z) d\Omega_n 
    - \frac{1}{(2\pi i)^2} \oint_{Q_\infty^-} 
        f_m^-(z) d\Omega_n. 
\eeqn
In particular, the second derivatives  
$\rd^2\calF/\rd a_j\rd a_k$ are identical to 
the matrix elements of the period matrix $\calT$ 
for the deformed curve $C(\vec{a},\vec{T})$. 
Moreover, by Riemann's bilinear relation, 
the mixed derivatives can also be written 
\beqn
    \frac{\rd^2 \calF}{\rd a_j \rd T_n} = 
    - \frac{1}{2\pi i}\oint_{\beta_j} d\Omega_n. 
\eeqn

Gorsky et al. \cite{bib:GMMM98} further proceeded 
to calculating these period integrals explicitly. 
This yields a formula for the second derivatives 
$\rd^2\calF/\rd T_m \rd T_n$ ($m,n = 1,\cdots,\ell$)  
in terms of a theta function, which is related to 
contact terms of topologically twisted $N = 2$ 
theories.  We shall return to this issue later on.

\subsection{Solutions for other classical gauge groups} 

A new feature that arises in the orthogonal and 
symplectic gauge groups is the ``parity'': The 
differentials and cycles in the Seiberg-Witten 
geometry have to respect the parity under the action 
of the involutions.  This is also the case for 
Whitham deformations.   Apart from this feature, 
it is straightforward to generalize the method 
of Gorsky et al. \cite{bib:GMMM98} to the other 
classical gauge groups, which we now present below 
\cite{bib:Takasaki99}. 

The first step of the construction is to seek for 
suitable meromorphic differentials of the form 
\beqn
    d\Omegahat_n = R_n(x) \frac{dz}{z}, \quad 
    R_n(x) = \mbox{polynomial}, 
\eeqn
with the following properties: 
\begin{itemize}
\item 
$R_n(x)$ is an odd polynomial. 
\item 
$(\rd/\rd u_j)d\Omegahat_n |_{z={\mathrm{const.}}}$ 
are holomorphic differentials on $C$. 
\end{itemize}
A solution to this problem can be found by applying 
the fractional power construction to $W(x)$: 
\beqn
    SO(2\ell+1) &:& 
    R_n(x) = \left( W(x)^{(2n-1)/(2\ell-1)} \right)_+, 
    \nonumber \\
    Sp(2\ell) &:& 
    R_n(x) = \left( W(x)^{(2n-1)/(2\ell+2)} \right)_+, 
    \nonumber \\
    SO(2\ell) &:& 
    R_n(x) = \left( W(x)^{(2n-1)/(2\ell-2)} \right)_+. 
\eeqn
Note that $R_n(x)$ is a polynomial of the form 
$x^n + \cdots$. In particular, as in the case of 
$SU(\ell+1)$, $d\Omegahat_1$ is nothing but the 
Seiberg-Witten differential $dS_\SW$.  

The other part of the construction is fully parallel 
to the case of $SU(\ell+1)$.   We have only to pay an 
extra attention to the parity.  It is easy to see that 
the differentials $dS$, $d\Omegahat_n$, $d\Omega_n$ 
and $d\omega_j$ are ``odd'' under the action of the 
involution of $\sigma_2$.  

The involution $\sigma'$ and the quotient curve 
$C' = C/\sigma'$ lead to an alternative view. 
The differentials $dS$, $d\Omegahat_n$, $d\Omega_n$ 
and $d\omega_j$ are all invariant under the involution 
$\sigma'$. Accordingly, they actually descend to (or, 
equivalently, are the pull-back of) differentials on 
$C' = C/\sigma'$.  

The status of the meromorphic differentials $d\Omega_n$ 
is particularly interesting from the second point of 
view.  These meromorphic differentials have two poles 
at $P_\infty^\pm$.  Since these two points are mapped 
to the same point $Q_\infty$ ($\xi = \infty$) of $C'$, 
the corresponding meromorphic differentials on $C'$ 
have a single pole at $Q_\infty$, and by a direct 
calculation, one can see that this is a pole of order 
$2n$.  As already mentioned, this is a property shared 
by the meromorphic differentials that arise in the KdV 
hierarchy.  More precisely, the singular behavior of 
those meromorphic differentials at $Q_\infty$ is such 
that 
\beqn
    d\Omega_n^\KdV = d\xi^{n-1/2} + \mbox{non-singular}. 
\eeqn
Our meromorphic differentials $d\Omega_n$ are a linear 
combination of those ``normalized'' meromorphic 
differentials.

\section{Application to Topologically Twisted 
$N = 2$ Theories} 
\setcounter{equation}{0}

\subsection{Tau functions and modular transformations}

The algebro-geometric tau functions of integrable 
hierarchies (KP, Toda, etc.) are determined by 
a set of algebro-geometric data (the so called 
Krichever data) including a non-singular complex 
algebraic curve (Riemann surface) $C$ of genus $g$   
\cite{bib:Krichever78,bib:Krichever77,bib:Dubrovin81,
bib:Mulase84,bib:Shiota86}.  Such a tau function 
can be generally written 
\beqn
    \tau(\vec{t}) = e^{2\pi iQ(\vec{t})}  
      \Theta(\sum_n t_n V_n + c \mid \calT),
    \quad 
    \vec{t} = (t_1,t_2,\cdots). 
\eeqn
Here $Q(\vec{t})$ is a quadratic form (including 
linear and constant terms), 
\beqn
    Q(\vec{t}) = 
        \frac{1}{2} \sum_{m,n} q_{mn} t_m t_n 
      + \sum_n r_n t_n + r_0, 
\eeqn
$V_n = (V^{(n)}_j)_{j=1,\cdots,g}$ and 
$c = (c_j)_{j=1,\cdots,g}$ are $g$-dimensional 
vectors, and $\Theta(w \mid \calT)$ denotes 
the Riemann theta function 
\beqn
    \Theta(w \mid \calT) = \sum_{n\in\bbZ^g}
      \exp\Bigl( \pi i n \cdot \calT n  
        + 2\pi i n \cdot w \Bigr), 
\eeqn
where the ``$\cdot$'' means the inner product, 
e.g., $n \cdot w = \sum_{j=1}^g n_j w_j$. 
The constants $r_n$, $r_0$ and $c_j$ are arbitrary, 
but this is not the case for $q_{mn}$ and $V^{(n)}_j$. 
As we illustrate below, they are given by some period 
integrals on $C$.

\subsubsection*{Example related to KP hierarchy} 
Let us consider the algebro-geometric tau functions 
of the KP hierarchy in a somewhat non-standard 
(in the sense specified below) formulation.  The 
algebro-geometric data are a non-singular algebraic 
curve $C$ of genus $g$ with a marked point $Q_\infty$, 
a local coordinate $\kappa$ in a neighborhood of 
$Q_\infty$ with $\kappa(Q_\infty) = 0$, a set of 
polynomials $f_n(\kappa)$ ($n = 1,2,\cdots$) in 
$\kappa^{-1}$ with constant coefficients, and 
a symplectic basis $\alpha_j,\beta_j$ 
($j = 1,\cdots,g$) of cycles on $C$.  
(This is the setup that we encounter when 
the Seiberg-Witten geometry for orthogonal 
and symplectic gauge groups is reformulated 
in the language of the quotient curve $C'$.) 
Let $d\omega_j$ ($j = 1,\cdots,g$) 
be a basis of holomorphic differentials on $C$ 
normalized by the condition 
\beqn
    \frac{1}{2\pi i} \oint_{\alpha_j} d\omega_k 
    = \delta_{jk}. 
\eeqn
The period matrix 
$\calT = (\calT_{jk})_{j,k=1,\cdots,g}$ is defined 
by the period integrals 
\beqn
    \frac{1}{2\pi i} \oint_{\beta_j} d\omega_k 
    = \calT_{jk}. 
\eeqn
Furthermore, a set of meromorphic differentials 
$d\Omega_n$ ($n = 1,2,\cdots$) are uniquely 
determined by the following conditions: 
\begin{itemize}
\item 
$d\Omega_n$ is non-singular outside $Q_\infty$, 
has a pole at $Q_\infty$, and the leading part 
of singularity at $Q_\infty$ is given by 
$df_n(\kappa)$, 
\beqn
    d\Omega_n = df_n(\kappa) + \mbox{non-singular}. 
\eeqn
\item 
The $\alpha$-periods of $d\Omega_n$ vanish, 
\beqn
    \oint_{\alpha_j} d\Omega_n = 0 \quad 
    (j = 1,\cdots,g). 
\eeqn
\end{itemize}
Now the constants $q_{mn}$ and $V^{(m)}_j$ are given 
by contour integrals along a small circle around 
$Q_\infty$: 
\beqn
    q_{mn} &=& - \frac{1}{(2\pi i)^2} 
      \oint_{Q_\infty} f_m(\kappa) d\Omega_n, 
    \nonumber \\
    V^{(m)}_j &=& - \frac{1}{(2\pi i)^2} 
      \oint_{Q_\infty} f_m(\kappa) d\omega_j, 
\eeqn
By Riemann's identity, one can readily see that 
$q_{mn}$ is symmetric and $V^{(m)}_j$ can be 
rewritten 
\beqn
    V^{(m)}_j = 
    \frac{1}{2\pi i} \oint_{\beta_j} d\Omega_m. 
\eeqn

\subsubsection*{Remarks} 
\begin{enumerate}
\item 
In the ``standard'' formulation of the KP hierarchy, 
$f_n(\kappa)$ is chosen to be $\kappa^{-n}$, so that 
the condition on the singular behavior of $d\Omega_n$ 
at $Q_\infty$ becomes 
\beqn
    d\Omega_n = d\kappa^{-n} + \mbox{non-singular}. 
\eeqn
Accordingly, the definition of $q_{mn}$ and $V^{(m)}_j$ 
has to be modified as 
\beqn
    q_{mn} &=& - \frac{1}{(2\pi i)^2} 
      \oint_{Q_\infty} \kappa^{-m} d\Omega_n, 
    \nonumber \\
    V^{(m)}_j &=& - \frac{1}{(2\pi i)^2} 
      \oint_{Q_\infty} \kappa^{-m} d\omega_j. 
\eeqn
but the other part remains intact.  The aforementioned 
formulation is simply to take an arbitrary set of 
directional vectors for the time variables $t_n$ 
in this standard formulation of the KP hierarchy. 
\item
In particular, suppose that if $C$ is a hyperelliptic 
curve of the form 
\beqn
    \eta^2 = R(\xi) = 
    \xi^{2g+1} + c_1 \xi^{2g} + \cdots + c_{2g+1}, 
\eeqn
$Q_\infty$ the point at infinity $\xi = \infty$, and 
$\kappa$ is given by 
\beqn
    \kappa = \xi^{-1/2}. 
\eeqn
Then in the standard formulation as mentioned above, 
all even members $d\Omega_{2n}$ of the meromorphic 
differentials become exact, 
\beqn
    d\Omega_{2n} = d\xi^n. 
\eeqn
thereby the directional vectors $V_{2n}$ for the even 
time variables $t_{2n}$  all vanish.  The coefficients 
$q_{mn}$ of the Gaussian factor also vanish for even 
indices.  This means that all the ``even'' flows become 
trivial, so that the KP hierarchy reduces to the KdV 
hierarchy.  
\end{enumerate}

\subsubsection*{Example related to Toda hierarchy} 
An immediate generalization of the KP-like setup 
is such that the algebraic curve has two marked 
points $P_\infty^\pm$.  Suppose that a local 
coordinate $\kappa_\pm$ at $P_\infty^\pm$ 
(with $\kappa_\pm(P_\infty^\pm) = 0$) and a 
symplectic basis $\alpha_j,\beta_j$ of cycles 
are given.  This is exactly the setup for 
constructing an algebro-geometric tau function 
of the Toda hierarchy \cite{bib:Ue-Ta84} 
(which is a hierarchy obtained from the 
two-dimensional $SU(\infty)$ Toda field theory).  
A standard formulation is  to introduce two 
series of time variables $t_n^\pm$ 
($n = 1,2,\cdots$) associated with the two 
marked points $P_\infty^\pm$.  We now consider 
a suitable linear combinations $t_n$ of those 
standard flows. 
(This is indeed the setup that takes place 
in the Seiberg-Witten geometry for $SU(\ell+1)$ 
gauge groups.)  
This linear combination is specified by 
a set polynomials $f_n^\pm(\kappa_\pm)$ in 
$\kappa_\pm^{-1}$ with constant coefficients.  
Given these data,   one can define meromorphic 
differentials $d\Omega_n$ by the following 
conditions: 
\begin{itemize}
\item 
$d\Omega_n$ is non-singular outside $P_\infty^\pm$, 
has two poles at $P_\infty^\pm$, and the leading 
part of singularity at $P_\infty^\pm$ is given 
by $df_n(\kappa_\pm)$, 
\beqn
    d\Omega_n = df_n^\pm(\kappa_\pm) 
      + \mbox{non-singular}. 
\eeqn
\item 
The $\alpha$-periods of $d\Omega_n$ vanish, 
\beqn
    \oint_{\alpha_j} d\Omega_n = 0 \quad 
    (j = 1,\cdots,g). 
\eeqn
\end{itemize}
Defining $q_{mn}$ and $V^{(m)}_j$ as 
\beqn
    q_{mn} &=& 
    - \frac{1}{(2\pi i)^2} \oint_{P_\infty^+} 
        f_m^+(\kappa_+) d\Omega_n 
    - \frac{1}{(2\pi i)^2} \oint_{P_\infty^-} 
        f_m^-(\kappa_-) d\Omega_n, 
    \nonumber \\ 
    V^{(m)}_j &=& 
    - \frac{1}{(2\pi i)^2} \oint_{P_\infty^+} 
        f_m^+(\kappa_+) d\omega_j 
    - \frac{1}{(2\pi i)^2} \oint_{P_\infty^-} 
        f_m^-(\kappa_-) d\omega_j, 
\eeqn
one obtains the tau function.  Also here, Riemann's 
bilinear identity implies that $q_{mn}$ is symmetric 
and that $V^{(m)}_j$ can be rewritten 
\beqn
    V^{(m)}_j = 
      \frac{1}{2\pi i} \oint_{\beta_j} d\Omega_m.  
\eeqn

We here briefly mention how these stuff are related 
to the Whitham deformations.  Upon turning on 
the Whitham deformations, $q_{mn}$ and $V^{(n)}_j$ 
are deformed to $\vec{T}$-dependent quantities: 
$q_{mn} \mapsto q_{mn}(\vec{T})$, 
$V^{(n)}_j \mapsto V^{(n)}_j(\vec{T})$. Comparing 
the definition of $q_{mn}$ and $V^{(n)}_j$ with 
the period integral formulae of the second 
derivatives of the deformed prepotential 
$\calF(\vec{T})$, one will soon find that 
\beqn
    q_{mn}(\vec{T}) = 
    \frac{\rd^2 \calF(\vec{T})}{\rd T_m \rd T_n}, 
    \quad 
    V^{(n)}_j(\vec{T}) = 
    \frac{\rd^2 \calF(\vec{T})}{\rd a_j \rd T_n}. 
\eeqn
These relations are also crucial in understanding 
the role of the Whitham deformations and the tau 
functions in topologically twisted $N = 2$ theories.  

We now turn to the modular property of these 
tau functions and its building blocks. Actually, 
this issue was studied in the eighties in the 
context of free fermions on a Riemann surface 
\cite{bib:AG-Mo-Va86,bib:KNTY88}. 
The modular property turns out to be model-independent, 
i.e., apply to the aforementioned general setup without 
specifying the integrable system, the algebro-geometric 
data, etc. \cite{bib:Takasaki98} 

The first step is to determine the transformations of 
the building blocks of the tau functions under the 
symplectic transformations 
\beqn
    \begin{array}{ll}
    \beta_j \mapsto A_{jk} \beta_k + B_{jk} \alpha_k, \\
    \alpha_j \mapsto C_{jk} \beta_k + D_{jk} \alpha_k, 
    \end{array}
    \quad 
    \left( \begin{array}{ll}
      A & B \\
      C & D
    \end{array} \right)  \in Sp(2g, \bbZ) 
\eeqn
of cycles.  This induces the well known transformation 
\beqn
    \calT \mapsto (A\calT + B) (C\calT + D)^{-1} 
\eeqn
of the period matrix $\calT$, which stands for the period 
matrix $\calP$ of $\Jac(C')$ if we consider the case of 
orthogonal and symplectic gauge groups.  The normalized 
holomorphic differentials $d\omega_j$ and the meromorphic 
differentials  $d\Omega_n$ transform as 
\beqn
    d\omega_j &\mapsto& [(C\calT + D)^{-1}]_{kj} d\omega_k, 
    \nonumber \\
    d\Omega_n &\mapsto& 
    d\Omega_n - \frac{1}{2\pi i} [(C\calT + D)^{-1} C]_{kj} 
    \cdot \oint_{\beta_j} d\Omega_n \cdot d\omega_k. 
\eeqn
Accordingly, $q_{mn}$ and $V^{(n)}_n$ transform as 
\beqn
    V^{(n)}_j &\mapsto& [(C\calT + D)^{-1}]_{kj} V^{(n)}_k, 
    \nonumber \\
    q_{mn} &\mapsto& 
    q_{mn} - [(C\calT + D)^{-1} C]_{kj} V^{(m)}_j V^{(n)}_k. 
\eeqn
It should be noted that these transformations already 
signal a possible connection with the contact terms in 
the $u$-plane integral of topological theories, which 
are known to obey substantially the same transformations. 

The next steps is to combine these transformations with 
the modular property of theta functions.  To this end, 
we introduce the theta functions with characteristics, 
\beqn
    && \Theta[\gamma,\delta](w \mid \calT) 
    \nonumber \\
    &=& \sum_{n\in\bbZ^g} \exp \Bigl( 
            \pi i (n + \gamma) \cdot \calT (n + \gamma) 
          + 2\pi i (n + \gamma) \cdot (w + \delta) \Bigr). 
\eeqn
These theta functions are known to obey the following 
modular transformations \cite{bib:Mumford83}: 
\beqn
    && \Theta[\gamma,\delta]\Bigl( 
         {}^t(C\calT + D)^{-1} w \mid 
         (A\calT + B)(C\calT + D)^{-1} \Bigr) 
    \nonumber \\
    &=& \epsilon \det(C\calT + D)^{1/2} 
        \exp(\pi i w \cdot (C\calT + D)^{-1}Cw)
        \Theta[\gamma',\delta'](w \mid \calT).  
\eeqn
Here $\epsilon$ is a eighth root of unity and 
$[\gamma',\delta']$  a transformed theta 
characteristic, both of which are determined 
by the symplectic matrix.  

We now examine the modular property of tau functions 
of the form 
\beqn
    \tau_{\gamma,\delta}(\vec{t}) = e^{2\pi iQ(\vec{t})} 
      \Theta[\gamma,\delta](\sum_n t_n V_n \mid \calT). 
    \quad 
\eeqn
Here (and in the following), $Q(\vec{t})$ is purely 
quadratic: 
\beqn
    Q(\vec{t}) = \frac{1}{2}\sum_{m,n}q_{mn}t_m t_n. 
\eeqn
As the identity 
\beqn
    \Theta[\gamma,\delta](w \mid \calT) = 
    e^{2\pi i\gamma\cdot(w+\delta)} 
      \Theta(w + \calT \gamma + \delta \mid \calT) 
\eeqn
implies, this is a special case of the aforementioned 
tau functions.  One may consider the more general 
tau functions, but these tau functions  
$\tau_{\gamma,\delta}(\vec{t})$ turn out to possess 
a better modular property.  From the modular 
transformations of the theta functions, indeed, we 
readily find that these tau functions transform as 
\beqn
    \tau_{\gamma,\delta}(\vec{t}) \mapsto 
    \epsilon \det(C\calT + D)^{1/2} 
    \tau_{\gamma',\delta'}(\vec{t}). 
\eeqn
Note that the multiplicative factor of the transformation 
is independent of $\vec{t}$; this is not the case for 
the more general tau function.  This fact, too, is 
a key to search for a connection with topological 
theories.

\subsection{$u$-plane integrals and contact terms in 
topologically twisted $N = 2$ theories} 

The topological twisting of an $N = 2$ supersymmetric 
gauge theory gives a topological field theory that detects 
the topology of the (compactified) four-dimensional 
space-time $X$ \cite{bib:Witten88}.  In the following, 
we consider the case where $X$ is simply connected.  The 
topological information is encoded in the correlators of 
observables, which are obtained by successively applying 
a descent operator $G$ to the Casimirs of the scalar 
field $\phi$ in the $N = 2$ vector multiplet, then 
integrating over a homology cycle of $X$.  

In the simply connected case, the relevant observables 
are supported by zero- and two-dimensional cycles.  
The correlators of those zero- and two-cycle observables 
can be collected into the generating function 
\beqn
    Z = \left< \exp\Bigl( \sum_j f_k \calO_k 
          + \sum_n g_n I_n(S) \Bigr) \right>, 
\eeqn
where $\calO_k$ and $I_n(S_n)$ are observables supported 
by a zero-cycle (point) $Q_k$ and a two-cycle (surface) 
$S_n$, and $f_k$ and $g_n$ are their coupling constants.  
Thus $\calO_k$ is just the value of a Casimir of $\phi$ 
at $Q_k$ (whose position itself is irrelevant in the 
correlators), and $I_n(S_n)$ is an integral of the form 
\beqn
    I_n(S_n) = \int_{S_n} G^2 \calP_n, 
\eeqn
where $\calP_n$ is yet another Casimir of $\phi$.  
In the case of the $SU(2)$ gauge group, these correlators 
give the Donaldson invariants of $X$ 
\cite{bib:Do-Kr90,bib:Fr-Mo91}. 

If, however, $X$ is a manifold with $b_2^+(X) = 1$ 
(e.g., complex rational surfaces), those observables 
loose topological invariance, and exhibit the so called 
``chamber structure'' or ``wall crossing phenomena''. 
Moore and Witten \cite{bib:Mo-Wi97} proposed a field 
theoretical interpretation of this phenomena, 
postulating that $Z$ is the sum 
\beqn
    Z = Z_D + Z_\calU 
\eeqn
of contributions of the strong coupling singularities 
and the bulk of the Coulomb moduli space (``$u$-plane'') 
$\calU$.  They determined an explicit form of $Z_\calU$ 
by examining the modular property of the (low energy 
effective) theory under the $Sp(2\ell,\bbZ)$ duality 
group ($\ell$ being the rank of the gauge group).  This 
is nothing but the group of symplectic transformations 
cycles in $C$ that we have discussed.  The outcome is 
an integral of the form 
\beqn
    Z_\calU = \int_\calU [dad\abar] A^\chi B^\sigma 
      \exp(U + \sum_{m,n} g_m g_n S_m \cdot S_n T_{mn}) 
      \Psi, 
\eeqn
where $A$ and $B$ are given by the formulae 
\beqn
    A = \alpha (\det \rd u_k/\rd a_j)^{1/2}, 
    \quad 
    B = \beta \Delta^{1/8},
\eeqn
$\alpha$ and $\beta$ are some constants, and 
$\Delta$ is essentially the discriminant of 
$W(x)^2 - 4\mu^2$;  $\chi$ and $\sigma$ are the 
Euler characteristic and the signature of $X$; 
$U$ is a collection of the effects of the zero-cycle 
observables; $S_m \cdot S_n$ is the intersection number 
of $S_m$ and $S_n$; $T_{mn}$ are the ``contact terms'' 
of the two-cycle observables; $\Psi$ is a lattice 
sum evaluating the photon partition function in the 
effective $U(1)^\ell$ theory.  

Our main concern lies in the contact terms.  According 
to Moore and Witten \cite{bib:Mo-Wi97} (for $SU(2)$) 
and Mari\~no and Moore \cite{bib:Ma-Mo97-98} (for 
other gauge groups), the contact terms are uniquely 
determined by the following properties: 
\begin{itemize}
\item 
Under the $Sp(2\ell,\bbZ)$ duality group, they transform 
as 
\beqn
    T_{mn} \mapsto T_{mn} - \frac{1}{4\pi i} 
      [(C\calT + D)^{-1} C]_{jk} 
      \frac{\rd \calP_m}{\rd a_j} 
      \frac{\rd \calP_n}{\rd a_k}.  
\eeqn
\item
In the semi-classical (i.e., weak coupling) limit as 
$\Lambda/a_j \to 0$, 
\beqn
    T_{mn} \to 0. 
\eeqn
\end{itemize}
Note that $\calP_n$ is understood to be a function on 
$\calU$.  Also recall that the matrix $\calT$ stands 
for the relevant $\ell$-dimensional Abelian variety 
--- the period matrix $\calT$ of $\Jac(C)$ in the 
case of $SU(\ell+1)$ gauge groups, and the period 
matrix $\calP$ of $\Jac(C')$ in the case of orthogonal 
and symplectic gauge groups. 

Remarkably, the modular transformations of the contact 
terms $T_{mn}$ are essentially of the same form as the 
modular transformations of $q_{mn}$.  (Note that 
$(C\calT + D)^{-1}C$ is a symmetric matrix.)   One might 
thus naively guess that $T_{mn}$ is identical to $q_{mn}$ 
(up to a multiplicative constant).  This comparison also 
suggests to identify $\rd \calP_n / \rd a_j$ with 
$V^{(n)}_j$.  This implies that, upon turning on the 
Whitham deformations, $\calP_n$ are given by 
\beqn
    \calP_n = {\mathrm{const.}} 
      \left.\frac{\rd \calF(\vec{T})}{\rd T_n}
      \right|_{\vec{T}=(1,0,0,\cdots)}, 
\eeqn
because of the relation 
$V^{(n)}_j = \rd^2 \calF(\vec{T})/\rd a_j \rd T_n$ 
in the Whitham deformations. 

A trouble in this naive identification is that $q_{mn}$ 
{\it do not} fulfill the correct semi-classical 
property mentioned above.\footnote{I overlooked this 
problem and wrongly identified $T_{mn}$ with $q_{mn}$ 
in my previous papers \cite{bib:Takasaki98,bib:Takasaki99}. 
I would like to take advantage of this opportunity 
to correct this error.} 
A correct identification, as pointed out by 
Gorsky et al.\cite{bib:GMMM98}, is achieved upon 
subtracting a singular part $q_{mn}^{\mathrm{sing}}$ 
from $q_{mn}$: 
\beqn
    T_{mn} = {\mathrm{const.}} 
      (q_{mn} - q_{mn}^{\mathrm{sing}}). 
\eeqn
Of course, the subtracted term $q_{mn}^{\mathrm{sing}}$ 
must be modular invariant in order to retain the 
aforementioned modular property.  Gorsky et al. 
worked out this separation for the $SU(\ell+1)$ 
theory by evaluating the period integrals for 
$q_{mn} = \rd^2 \calF/\rd T_m \rd T_n$, and 
obtained the formula 
\beqn
    q_{mn} - q_{mn}^{\mathrm{sing}} = 
    - \frac{2\beta^2}{(2\pi i)^2} 
      \frac{\rd \log \Theta_E(0 \mid \calT)}
           {\rd \calT_{jk}} 
      \frac{\rd \calH_{m+1}}{\rd a_j}
      \frac{\rd \calH_{n+1}}{\rd a_k}, 
\eeqn
where $\beta = 2\ell + 2$, $\Theta_E$ is a theta function 
with an ``even'' half-characteristic, and $\calH_{n+1}$ 
is defined by the residue 
\beqn
    \calH_{n+1} = \frac{\ell+1}{n} 
      \Res_{x=\infty} P(x)^{n/(\ell+1)}dx. 
\eeqn
Thus in order to complete the identification, we have 
to interpret $\calP_n$ as $\calH_{n+1}$ (up to 
a multiplicative constant).  The singular part 
$q_{mn}^{\mathrm{sing}}$ itself is defined by 
\beqn
    q_{mn}^{\mathrm{sing}} = 
    - \frac{\beta}{2\pi i} \calH_{m+1,n+1}, 
\eeqn
where 
\beqn
    \calH_{m+1,n+1} = \frac{\ell+1}{mn} 
      \Res_{x=\infty} P(x)^{m/(\ell+1)}
        d(P(x)^{n/(\ell+1)})_+. 
\eeqn
It should be also noted that these calculations 
are valid for the range of $m,n \le \ell$ only.

\subsection{Blowup formula and tau functions} 

Another approach to the evaluation of contact terms 
is based on the blowup formula.  The aforementioned 
theta formula of contact terms was indeed first 
derived by Losev et al. \cite{bib:Lo-Ne-Sh97-98} 
using the technique of blowup.  The blowup formula 
was reformulated by Moore and Witten \cite{bib:Mo-Wi97} 
and by Mari\~no and Moore \cite{bib:Ma-Mo97-98} 
in the language of the $u$-plane integral.  This 
reveals a close connection to the tau functions 
$\tau_{\gamma,\delta}(\vec{t})$.  In particular, 
as Marino pointed out recently \cite{bib:Marino99}, 
the theta function in the aforementioned formula of 
contact terms is actually the same as the theta 
function in the tau function.   We present an 
outline of these observations below.  

The blowup formula relates the topological 
correlators between $X$, which is now assumed 
to be a complex algebraic surface with $b_2^+(X) = 1$, 
and its blowup $\Xhat$ at a point $Q$ of $X$.  
The inverse image of $Q$ in the blowup map 
$\Xhat \to X$ is a rational curve (called 
an ``exceptional curve'') with self-intersection 
$-1$.  Let $B$ denote its homology class in 
$H_2(\Xhat,\bbZ)$.  The effect of blowup on 
the homology is simply to add to $H_2(X,\bbZ)$ 
a rank one module generated by $B$: 
\beqn
    H_2(\Xhat,\bbZ) = H_2(X,\bbZ) \oplus \bbZ B. 
\eeqn
One can now insert a set of observables $I_n(B)$ 
supported by $B$ into the generating function on 
$\Xhat$: 
\beqn
   Z_\Xhat = \left< \exp\Bigl( \sum_k f_k \calO_k 
     + \sum_n g_n I_n(S) + \sum_n t_n I_n(B) \Bigr)  
     \right>.  
\eeqn
According to the formulation of Mari\~no and Moore 
\cite{bib:Ma-Mo97-98}, the effect of blowup in 
the $u$-plane integral is to modify the integrand 
for $X$ as follows: 
\begin{enumerate}
\item 
In accordance with the decomposition of $H_2$ mentioned 
above, the photon partition function $\Psi_\Xhat$ 
becomes a product of $\Psi_X$ and a theta function 
with an even half-characteristic: 
\beqn
    \Psi_\Xhat = 
    \Theta[0,\delta](\sum_n t_n V_n \mid \calT) \Psi_X, 
    \quad \delta = (1/2,\cdots,1/2). 
\eeqn
\item 
The exponential function containing contact terms 
undergoes a new contribution from the observables 
supported by $B$, which appear as the multiplicative 
factor 
\beqn
    \exp(- \sum_{m,n} T_{mn} t_m t_n). 
\eeqn
The negative sign in the exponent stems from the 
self-intersection $B \cdot B = -1$ of the exceptional 
divisor. 
\item 
The Euler characteristic and the signature 
change as: $\chi(\Xhat) = \chi(X) + 1$, 
$\sigma(\Xhat) = \sigma(X) - 1$. 
The measure factor $A^\chi B^\sigma$ is thereby 
multiplied by 
\beqn
    AB^{-1} = \det(\rd u_k/\rd a_j)^{1/2} \Delta^{-1/8}. 
\eeqn
\end{enumerate}

Let us consider the product of the first two 
factors (i.e., the Gaussian and the theta 
function), which we call $\tauhat_{0,\delta}(\vec{t})$: 
\beqn
    \tauhat_{0,\delta}(\vec{t}) = 
    \exp(- \sum_{m,n} T_{mn} t_m t_n) 
    \Theta[0,\delta](\sum_n t_n V_n \mid \calT).  
\eeqn
Mari\~no and Moore \cite{bib:Ma-Mo97-98} remarked 
that this is essentially the tau function of the 
Toda system.  More precisely, as we have mentioned, 
this differs from the true tau function 
$\tau_{0,\delta}(\vec{t})$ by a Gaussian factor 
$\exp(2\pi iQ_{\mathrm{sing}}(\vec{t}))$, where 
$Q_{\mathrm{sing}}(\vec{t})$ is the quadratic 
form with coefficients $q_{mn}^{\mathrm{sing}}$.  
Thus we have 
\beqn
    \tauhat_{0,\delta}(\vec{t}) 
    &=& \exp(- Q_{\mathrm{sing}}(\vec{t})) 
       \tau_{0,\delta}(\vec{t}) 
  \nonumber \\
    &=& \exp(2\pi i(Q(\vec{t}) 
          - Q_{\mathrm{sing}}(\vec{t})) 
        \Theta[0,\delta](\sum_n t_n V_n \mid \calT). 
\eeqn
Since the subtracted term $Q_{\mathrm{sing}}(\vec{t})$ 
is modular invariant, this modification does not spoil 
the modular transformation of $\tau_{0,\delta}(\vec{t})$; 
the subtraction is necessary for the correct semi-classical 
behavior of the integrand of the $u$-plane integral.  

Moreover, according to Mari\~no \cite{bib:Marino99}, 
the third factor in the blowup process is nothing 
but the inverse of the zero-value of 
$\tauhat_{0,\delta}(\vec{t})$: 
\beqn
    A B^{-1} = \frac{1}{\tauhat_{0,\delta}(\vec{0})}, 
    \quad \vec{0} = (0,0,\cdots). 
\eeqn
The product of the three factor thus boils down to 
the tau quotient 
\beqn
    \frac{\tauhat_{0,\delta}(\vec{t})}
         {\tauhat_{0,\delta}(\vec{0})} = 
    \frac{\tau_{0,\delta}(\vec{t})}
         {\tau_{0,\delta}(\vec{0})}. 
\eeqn
Since $\tau_{0,\delta}(\vec{t})$ and 
$\tau_{0,\delta}(\vec{0})$ have the same modular 
property, the quotient is modular invariant --- 
a property to be required for the consistency 
of the $u$-plane integral. Furthermore, this 
quotient has to be non-singular in the 
semi-classical region of the $u$-plane.  
On the basis of these requirements, Mari\~no 
\cite{bib:Marino99} eventually derives the theta 
formula 
\beqn
    T_{mn} = 
    2\pi i 
    \frac{\rd\log\Theta[0,\delta](0 \mid \calT)}
         {\rd \calT_{jk}}
    V^{(m)}_j V^{(n)}_k, 
\eeqn
thus reproducing the theta formula of Gorsky et al. 
\cite{bib:GMMM98} from an entirely different route. 
This result also also shows that the theta function 
$\Theta_E$ in the formula of Gorsky et al. is 
actually given by $\Theta[0,\delta]$.

\section{Conclusion}
\setcounter{equation}{0}

We have seen some new aspects of integrable 
systems discovered in the recent studies of $N = 2$ 
supersymmetric gauge theories and the topologically 
twisted versions.  A particularly impressive lesson 
is that the combination of Whitham deformations 
and tau functions can be a surprisingly powerful 
tool.  This fact is demonstrated in Mari\~no's 
beautiful exposition \cite{bib:Marino99}, part of 
which we have reviewed in the last section.  

A number of problems still remain to be addressed.  
A central issue will be to extend the present 
approach based on integrable systems to other 
cases, such as the theories with matter multiplets, 
exceptional gauge groups, etc. Contact terms and 
the blowup formula for theories with matter 
multiplets have been studied to some extent 
\cite{bib:Mo-Wi97,bib:Lo-Ne-Sh97-98,bib:Ma-Mo97-98}. 
Perhaps the most intriguing are the theories with 
an adjoint matter multiplet (equivalently, the mass 
deformed $N = 4$ theories).  The blowup formula in 
the topological versions of these theories, too, will 
be described by some tau function (of an elliptic
Calogero-Moser system?).  Another interesting case 
can be found in toroidally compactified tensionless 
strings (also called $E$-strings) and related 
$N = 2$ or mass deformed $N = 4$ gauge theories 
\cite{bib:Ganor96,bib:Ga-Mo-Se96,bib:Le-Ma-Wa96,
bib:Mi-Ne-Wa97,bib:MNCW98}. 
The $E_8$ theta function $\Theta_{E_8}$ arising 
in those models will be connected with the tau 
function of a yet unidentified integrable system 
underlying the Seiberg-Witten curves of those 
theories.

\section*{Acknowledgements}

I am grateful to Toshio Nakatsu, Masahiko Saito, 
Yuji Shimizu and Yasuhiko Yamada for useful 
discussions. This work is partly supported by the 
Grant-in-Aid for Scientific Research (No. 10640165) 
of the Ministry of Education, Science and Culture. 


\end{document}